\documentclass[pre,aps,floatfix,twocolumn, superscriptaddress]{revtex4-1}

\usepackage{amsmath}

\usepackage{amsfonts}

\usepackage{amssymb}
\usepackage{color,soul}
\usepackage{graphicx}
\usepackage{empheq}
\usepackage{verbatim}

\usepackage{subfigure}

\usepackage{lipsum}

\newcommand{\vc}{\mathbf}

\begin{document}

\title{Transport formulas for multi-component plasmas within the effective potential theory framework}

\author{Grigory~Kagan}
\email[E-mail: ]{kagan@lanl.gov}
\affiliation{Theoretical Division, Los Alamos National Laboratory, Los Alamos, NM 87545}
\author{Scott D. Baalrud}
\affiliation{Department of Physics and Astronomy, University of Iowa, Iowa City, IA 52242}

\date{\today}

\begin{abstract}

The recently proposed effective potential theory [Phys. Rev. Lett. 110, 235001 (2013)] allows evaluating transport in coupled plasmas with the well-developed  formalisms for systems with binary collisions.  To facilitate practical implementation of this concept in fluid models of multi-component plasmas, compact expressions for the transport coefficients in terms the generalized Coulomb logarithms are summarized from existing prescriptions. For weakly coupled plasmas, characterized by Debye-shielded Coulomb interaction potential, expressions become fully analytical. In coupled plasmas the generalized Coulomb logarithms need to be evaluated numerically. Routines implementing the described formalisms are included as supplemental material.

\end{abstract}

\maketitle


\section{Overview}

Evaluation of the transport coefficients for coupled plasmas is greatly complicated by the many-body physics of particle collisions. The recently proposed effective potential theory (EPT) addresses this issue by stipulating that, as far as transport is concerned,  collisions can be considered effectively binary even at finite coupling, with the many-body physics manifesting itself solely through modifying the interaction potential between the two colliding particles~\cite{baalrud2013effective}. In turn, the effective potential enters expressions for the transport coefficients through the so-called ``generalized Coulomb logarithms", which are closely related to the standard gas-kinetic cross sections. 

The resulting transport predictions for a one component plasma (OCP) prove in a remarkable agreement with molecular dynamics (MD) simulations, encouraging extension of the EPT concept to the case of a plasma with multiple ion species. While kinetic calculations for multi-component systems are  more complex, the problem is  well explored in the literature on diluted gas mixtures. In this Note we summarize the existing transport results in the form convenient for practical use.

Local transport formalisms for systems with binary collisions assume that the distribution function $ f_{\alpha}$ of a given species $\alpha$ weakly deviates from equilibrium, $\delta f_{\alpha} \equiv f_{\alpha} - f_{\alpha}^{(0)}  \ll f_{\alpha}^{(0)}$, due to the Knudsen number $N_K \equiv \lambda/L$ being small, where $\lambda$ and $L$ are the characteristic mean free path and background scale, respectively.  The linearized Boltzmann equation is solved for $\delta f_{\alpha}$ whose moments give transport coefficients of interest.

In the commonly used Chapman-Enskog approach, the solution for $\delta f_{\alpha}$ is obtained by expanding it over a set of orthogonal polynomials of the particle velocity $\vec{v}$. Accordingly, precision of the resulting transport coefficients is governed by the number $\xi$ of the so-called Sonyne polynomials kept in the expansion over the radial component of the velocity. Following earlier works we will denote the approximation level, in which transport quantity $Q$ is evaluated, by $[Q]_{\xi}$. 
Transport calculations based on Grad's method use different precision nomenclature, but it is straightforward to observe that the orthogonal polynomials employed there are the same and therefore local transport results  are identical to those obtained with the Chapman-Enskog approach. In particular, Grad's 21N results by Zhdanov~\cite{zhdanov2002transport}, in Chapman-Enskog's nomenclature correspond to $\xi = 3$ for the heat and diffusive fluxes and $\xi = 2$ for viscosity.

One difference between the  neutral gas mixtures and unmagnetized plasmas with multiple ion species is presence of electrons. Due to their small mass the energy exchange between them and ions  is much slower than equilibration within them or any of the ion species. Consequently, electron temperature should generally be distinguished in fluid plasma models. Also, in a vast range of scenarios plasmas are quasi-neutral and so, if $N$ were the total number of plasma species, there would be only $N-2$ independent species concentrations as opposed to $N-1$ in an $N$-component gas mixture. However, these issues can be easily circumvented by considering separately the two subsystems: all the ion species (ionic mixture) on the one hand and electrons on the other hand, which interact through collisions and fields. Evaluation of the ion transport then reduces to the classical problem of a mixture under external forces, making it possible to use the well-established prescriptions from the conventional kinetic theory of diluted gases.

Accordingly, in what follows we let $N$ denote the number of the ion species and exploit results from various sources obtained with either Chapman-Enskog~\cite{hirschfelder1954molecular, devoto1966transport, ferziger1972mathematical} or Grad~\cite{zhdanov2002transport} methods. The resulting compact representation for the transport coefficients is summarized in Sec.~\ref{app: formulary}. These formulas involve matrix elements, whose expressions in terms of the generalized Coulomb logarithms $\Xi_{\alpha\beta}^{(l,k)}$ are given in Sec.~\ref{app: matrix-elem}. Once the effective potential, and therefore $\Xi_{\alpha\beta}^{(l,k)}$, are known equations of Sections~\ref{app: formulary} and \ref{app: matrix-elem} provide explicit transport results. In particular, in the weakly coupled limit considered in Sections~\ref{app: matrix-elem-weak} and \ref{app: diff-weak}, $\Xi_{\alpha\beta}^{(l,k)}$  can be calculated analytically, thereby giving fully analytical expressions for all the transport coefficients.
Finally, Section~\ref{sec: routines} describes the numerical routines, which implement the formalisms for weakly and arbitrarily coupled plasmas.

\section{Matrix representation for transport coefficients}
\label{app: formulary}

In what follows, $x_{\alpha} = n_{\alpha}/n_i$ and $c_{\alpha} = \rho_{\alpha}/\rho$ denote the number and mass fractions of the ion species $\alpha$, respectively, where $n_{\alpha}$ and $\rho_{\alpha}$ are the number and mass densities of the ion species $\alpha$, respectively, and $n_i = \sum_{\alpha} n_{\alpha}$ and $\rho=\sum_{\alpha} \rho_{\alpha}$ are the total number and mass densities of the ionic mixture, respectively. Partial pressure of the ion species $\alpha$ is denoted by $p_{\alpha}$ and the total pressure of the ionic mixture is denoted by $p_i = \sum_{\alpha} p_{\alpha}$. Finally, $m_{\alpha} $ and $Z_{\alpha} $ are the particle mass and charge number of the ion species $\alpha$, respectively.

We also define the collision frequency  between plasma species $\alpha$ and $\beta$ by
\begin{equation}
\label{eq: nu}
\nu_{\alpha \beta} = \frac{4 \sqrt{2\pi} Z_{\alpha}^2 Z_{\beta}^2 e^4 \gamma_{\alpha\beta}^{3/2} n_{\beta}}
{3   \mu_{\alpha\beta}^2 } 
\Xi_{\alpha \beta}.
\end{equation}
In  Eq.~(\ref{eq: nu}), $\mu_{\alpha\beta} = m_{\alpha} m_{\beta}/(m_{\alpha} + m_{\beta})$ is the reduced mass and $\gamma_{\alpha\beta} = \gamma_{\alpha} \gamma_{\beta}/(\gamma_{\alpha} + \gamma_{\beta})$ with $\gamma_{\alpha} \equiv m_{\alpha} /(k_B T_{\alpha})$  and $T_1 = T_2 \equiv T_i$ need to be set for ion species with comparable masses. Finally, $\Xi_{\alpha \beta} \equiv \Xi_{\alpha \beta}^{(1,1)}$ is the lowest order generalized Coulomb logarithm, which was introduced in Ref.~\cite{baalrud2013effective}. Equation~(\ref{eq: nu}) reduces to the familiar expression in the weakly coupled limit, in which $\Xi_{\alpha \beta}$ becomes the conventional Coulomb logarithm $\ln{\Lambda}$~\cite{baalrud2014extending}. 

\subsection{Diffusive flux}
\label{app: diff}

Diffusive velocity of the ion species $\alpha$ is given by
\begin{equation}
\label{eq: diff-flux-gen}
\vec{V}_{\alpha} = 
- \sum_{\beta} D_{\alpha\beta} \vec{d}_{\beta} 
+ D_{\alpha}^{(T)} \nabla \ln{T_i},
\end{equation}
where  $\vec{d}_{\alpha}$ includes all the thermodynamic forces other than $\nabla T_i$:
\begin{equation}
\label{eq: diff-force}
\vec{d}_{\alpha} = 
\frac{\nabla p_{\alpha} - c_{\alpha} \nabla p_i}{p_i} -
\frac{\rho_{\alpha}}{p_i} \Bigl(  \vec{X}_{\alpha} - \sum_{\beta = 1}^N  c_{\beta}  \vec{X}_{\beta}  \Bigr).
\end{equation}
In Eq.~(\ref{eq: diff-force}), $\vec{X}_{\beta}$ are  forces that are external with respect to the ionic mixture. In the absence of the magnetic field and  forces that are external with respect to the plasma as a whole, $\vec{X}_{\beta}$ only includes the thermal force $\vec{ R}_{\beta e}^{(T)}$ exerted by electrons on the ion species $\beta$ and the electric field $\vec{E}$:
\begin{equation}
\label{eq: ext-force}
\vec{X}_{\beta} = \vec{ R}_{\beta e}^{(T)}/\rho_{\beta} +  Z_{\beta} e \vec{E}/m_{\beta},
\end{equation}
since the effect of the electron-ion dynamic friction on the ion transport can be neglected for realistic electron currents~\cite{kagan2012electro}.

To evaluate the diffusive flux from the background gradients one thus needs to know the ordinary and thermal diffusion coefficients, $D_{\alpha\beta}$ and $D_{\alpha}^{(T)}$. 
A number of equivalent representations for these can be found in literature~\cite{hirschfelder1954molecular, devoto1966transport, ferziger1972mathematical, zhdanov2002transport}.  
Here we utilize the formalism by Ferziger and Kaper~\cite{ ferziger1972mathematical}   and use the Kramers rule to write the $\xi$th Chapman-Enskog approximation to the ordinary and thermo-diffusion coefficients in the form of Ref.~\cite{devoto1966transport}
\begin{align}
\nonumber
[D_{\alpha\beta}]_{\xi} &= 
- \frac{4}{25 n_i  |\tensor{M}|} \times  \\
 \label{eq: ordin-diff}
 &\begin{vmatrix}
  \tensor{M}^{(0,0)} & \tensor{M}^{(0,1)}  & \ldots   & \tensor{M}^{(0,\xi-1)} &\vec{\delta}_{k\beta} - \vec{c }_{k} \\
  \tensor{M}^{(1,0)} & \tensor{M}^{(1,1)}  & \ldots   & \tensor{M}^{(1,\xi-1)} & \vec{0}  \\
  \vdots                   & \vdots                    & \ddots  & \vdots                        & \vdots \\
  \tensor{M}^{(\xi-1,0)} & \tensor{M}^{(\xi-1,1)} & \ldots   & \tensor{M}^{(\xi-1,\xi-1)} & \vec{0}\\
 \vec{ \delta }_{k\alpha}        & \vec{0}                                & \ldots          &             \vec{0}        & 0
 \end{vmatrix}
\end{align}
and
\begin{multline}
\label{eq: thermo-diff}
[D_{\alpha}^{(T)}]_{\xi}=
- \frac{2}{5 n_i  |\tensor{M}|}  \times \\
 \begin{vmatrix}
  \tensor{M}^{(0,0)} & \tensor{M}^{(0,1)}  & \ldots   & \tensor{M}^{(0,\xi-1)} & \vec{0} \\
  \tensor{M}^{(1,0)} & \tensor{M}^{(1,1)}  & \ldots   & \tensor{M}^{(1,\xi-1)} & \vec{x}_{k}  \\
  \vdots                   & \vdots                    & \ddots  & \vdots                        & \vdots \\
  \tensor{M}^{(\xi-1,0)} & \tensor{M}^{(\xi-1,1)} & \ldots   & \tensor{M}^{(\xi-1,\xi-1)} & \vec{0}\\
  \vec{ \delta }_{k\alpha}        & \vec{0}                                & \ldots          &               \vec{0}        & 0 
 \end{vmatrix},
\end{multline}
where blocks $\tensor{M}^{(i,j)}$ are $N \times N$ matrices, whose elements are provided in the next subsection. In Eqs.~(\ref{eq: ordin-diff}) and (\ref{eq: thermo-diff}),  $|\tensor{M}|$ denotes the  determinant of the $\xi N \times \xi N$ matrix $\tensor{M}$ composed of $\tensor{M}^{(i,j)}$. The  determinants in the numerator are obtained by appending $\tensor{M}$ with a row and a column that are, in turn, composed of  $N$-element vectors indicated by the arrow sign and the last element, scalar 0. The $k$-th element in such a vector is given by the corresponding expressions, in which $\delta_{kl}$ is the Kronecker delta and $v_k$ appearing in the upper right corner in the numerator on the right side of Eq.~(\ref{eq: ordin-diff}) is equal to $0$ for $k=1$ and $\delta_{k\beta} - c_k$ for $2 \leq k \leq N$.

In the employed formalism the ordinary diffusion coefficients are symmetric, $D_{\alpha\beta} = D_{\beta\alpha}$, and also satisfy the constraints 
$\sum_{\alpha} c_{\alpha} D_{\alpha\beta} = 0$ for $\beta = 1...N$, so there are only $N(N-1)/2$ independent coefficients. Thermo-diffusion coefficients satisfy the constraint $\sum_{\alpha} c_{\alpha} D_{\alpha}^{(T)} = 0$, so there are $N-1$ independent coefficients. It can be observed straight from Eq.~(\ref{eq: thermo-diff}) that thermo-diffusion vanishes in the lowest order approximation.

Finally, we notice that the above expressions provide the kinetic part of the diffusive flux. At finite coupling there should generally be a thermodynamic prefactor.  In the presented formalism it is included  through the partial ionic pressure in the first term on the right side of Eq.~(\ref{eq: diff-force}) and can be retrieved from the equation of state of coupled multi-component plasma.


\subsection{Heat flux}

The heat flux of the ion species $\alpha$ is written as
\begin{equation}
\label{eq: heat-flux-single}
\vec{q}_{\alpha} = - \lambda'_{\alpha} \nabla T_i 
-n_i T_i D_{\alpha}^{(T)} \vec{d}_{\alpha} 
- h_{\alpha} \vec{V}_{\alpha} ,
\end{equation}
where $h_{\alpha}$ is the  enthalpy density of the species $\alpha$ and the heat conductivity $\lambda'_{\alpha}$ is given by
\begin{align}
\nonumber
[\lambda'_{\alpha}]&_{\xi}  =
- \frac{1}{|\tensor{M}|} \times  \\
\label{eq: heat-conduct-single}
& \begin{vmatrix}
  \tensor{M}^{(0,0)} & \tensor{M}^{(0,1)}  & \ldots   & \tensor{M}^{(0,\xi-1)} & \vec{0} \\
  \tensor{M}^{(1,0)} & \tensor{M}^{(1,1)}  & \ldots   & \tensor{M}^{(1,\xi-1)} & \vec{x_k \delta_{k\alpha} }  \\
  \vdots                   & \vdots                    & \ddots  & \vdots                        & \vdots \\
  \tensor{M}^{(\xi-1,0)} & \tensor{M}^{(\xi-1,1)} & \ldots   & \tensor{M}^{(\xi-1,\xi-1)} & \vec{0}\\
 \vec{0}        & \vec{x}_k                               & \vec{0}        ~~~~\ldots  &               \vec{0}        & 0 
 \end{vmatrix},
\end{align}
where the matrix $\tensor{M}$ and notation are the same as in the preceding subsection. The total ion heat flux is written as
\begin{equation}
\label{eq: heat-flux-total}
\vec{q}_{\alpha} = - \lambda'_{i} \nabla T_i 
-n_i T_i \sum_{\alpha} D_{\alpha}^{(T)} \vec{d}_{\alpha} 
- \sum_{\alpha} h_{\alpha} \vec{V}_{\alpha} ,
\end{equation}
where the total heat conductivity of the ionic mixture is given by
\begin{align}
\label{eq: heat-conduct-total}
[\lambda'_{i}]_{\xi}  &=
-\frac{1}{|\tensor{M}|} \times  \\
 &\begin{vmatrix}
  \tensor{M}^{(0,0)} & \tensor{M}^{(0,1)}  & \ldots   & \tensor{M}^{(0,\xi-1)} & \vec{0} \\
  \tensor{M}^{(1,0)} & \tensor{M}^{(1,1)}  & \ldots   & \tensor{M}^{(1,\xi-1)} & \vec{x}_k  \\
  \vdots                   & \vdots                    & \ddots  & \vdots                        & \vdots \\
  \tensor{M}^{(\xi-1,0)} & \tensor{M}^{(\xi-1,1)} & \ldots   & \tensor{M}^{(\xi-1,\xi-1)} & \vec{0}\\
 \vec{0}        & \vec{x}_k                               & \vec{0}        ~~~~\ldots  &               \vec{0}        & 0 
 \end{vmatrix}
\end{align}

\subsection{Viscosity}

The viscous stress tensor of the ion species $\alpha$ is written as $\tensor{\pi}_{\alpha} = 
- \eta_{\alpha} \tensor{W}$, where the rate-of-strain tensor $\tensor{W}$ is defined by
\begin{equation}
\label{eq: strain}
W_ {jk} = \frac{\partial u_{ j}}{\partial x_k} + \frac{\partial u_{ k}}{\partial x_j} - \frac{2}{3} \nabla\cdot \vec{u}
\end{equation}
with $\vec{u}$ being the plasma center-of-mass velocity. The partial viscosity coefficient is given by
\begin{multline}
\label{eq: partial-visc}
[\eta_{\alpha}]_{\xi}=
- \frac{1}{|\tensor{L}|}  \times \\
 \begin{vmatrix}
  \tensor{L}^{(0,0)} & \tensor{L}^{(0,1)}  & \ldots   & \tensor{L}^{(0,\xi-1)} & \vec{x_k \delta_{k\alpha} }  \\
  \tensor{L}^{(1,0)} & \tensor{L}^{(1,1)}  & \ldots   & \tensor{L}^{(1,\xi-1)} &  \vec{0}   \\
  \vdots                   & \vdots                    & \ddots  & \vdots                        & \vdots \\
  \tensor{L}^{(\xi-1,0)} & \tensor{L}^{(\xi-1,1)} & \ldots   & \tensor{L}^{(\xi-1,\xi-1)} &  \vec{0} \\
 \vec{x}_k         & \vec{0}                              & \ldots          &                \vec{0}         & 0 
 \end{vmatrix},
\end{multline}
and the total viscosity of the ionic mixture is given by
\begin{multline}
\label{eq: total-visc}
[\eta_{i}]_{\xi}=
- \frac{1}{|\tensor{L}|}  \times \\
 \begin{vmatrix}
  \tensor{L}^{(0,0)} & \tensor{L}^{(0,1)}  & \ldots   & \tensor{L}^{(0,\xi-1)} & \vec{x}_k  \\
  \tensor{L}^{(1,0)} & \tensor{L}^{(1,1)}  & \ldots   & \tensor{L}^{(1,\xi-1)} &  \vec{0}   \\
  \vdots                   & \vdots                    & \ddots  & \vdots                        & \vdots \\
  \tensor{L}^{(\xi-1,0)} & \tensor{L}^{(\xi-1,1)} & \ldots   & \tensor{L}^{(\xi-1,\xi-1)} &  \vec{0} \\
 \vec{x}_k         & \vec{0}                              & \ldots          &                \vec{0}         & 0 
 \end{vmatrix}.
\end{multline}

\begin{widetext}
\section{Matrix elements in terms of the generalized Coulomb logarithms}
\label{app: matrix-elem}

It is convenient to introduce $ \bar{\Xi}_{\alpha \beta}^{(l,k)} \equiv \Xi_{\alpha \beta}^{(l,k)}/\Xi_{\alpha \beta}^{(1,1)}$. Then, elements of matrix $\tensor{M}$ can be written as follows: for the first row of the uppermost leftmost block ($i=j=0,\alpha=1$)
\begin{equation}
\label{eq: M-elem-1}
M_{1\beta}^{(0,0)} = c_{\beta},~ \beta =1..N,
\end{equation}
for the first rows of the remaining uppermost blocks ($i=0, 0<j\leq \xi-1,\alpha=1$)
\begin{equation}
\label{eq: M-elem-1}
M_{1\beta}^{(0,j)} = 0,~ \beta =1..N
\end{equation}
and for all other elements
\begin{equation}
\label{eq: M-elem-2}
M_{\alpha\beta}^{(i,j)} = \frac{8(m_{\alpha}m_{\beta})^{1/2}}{75 T_i }
\Bigl( \delta_{\alpha\beta} \sum_{\chi =1}^{N} x_{\alpha} x_{\chi} A_{\alpha\chi}^{(i,j)} +
x_{\alpha} x_{\beta} B_{\alpha\beta}^{(i,j)} \Bigr),
\end{equation}
where $A_{\alpha\beta}^{(i,j)}$ and $B_{\alpha\beta}^{(i,j)}$ are related to standard bracket integrals~\cite{ ferziger1972mathematical}  
so one can find $A_{\alpha \beta}^{(l,k)} = 3  \nu_{\alpha \beta}/(16  n_{\beta}) \bar{A}_{\alpha \beta}^{(l,k)} $ and $B_{\alpha \beta}^{(l,k)} = 3 \nu_{\alpha \beta}/(16  n_{\beta}) \bar{B}_{\alpha \beta}^{(l,k)} $ with $\nu_{\alpha \beta}$ defined by Eq.~(\ref{eq: nu}) and
\begingroup
\allowdisplaybreaks
\begin{align}
\label{eq: bracket-1}
\bar{A}_{\alpha\beta}^{(0,0)} &= 8 \mu_{\beta}  \\
\bar{A}_{\alpha\beta}^{(0,1)} &= 8 \mu_{\beta}^2  \Bigl(  \frac{5}{2}   -  
\bar{\Xi}_{\alpha \beta}^{(1,2)} \Bigr)\\
\bar{A}_{\alpha\beta}^{(1,1)} &= 8 \mu_{\beta}  \Bigl[\frac{5}{4} (6 \mu_{\alpha}^2 + 5 \mu_{\beta}^2)   -  
5 \mu_{\beta}^2 \bar{\Xi}_{\alpha \beta}^{(1,2)} +
\mu_{\beta}^2 \bar{\Xi}_{\alpha \beta}^{(1,3)} +
2 \mu_{\alpha} \mu_{\beta} \bar{\Xi}_{\alpha \beta}^{(2,2)} \Bigr]\\
\bar{A}_{\alpha\beta}^{(0,2)} &= 4 \mu_{\beta}^3  \Bigl(  \frac{35}{4} -  
7 \bar{\Xi}_{\alpha \beta}^{(1,2)} +
\bar{\Xi}_{\alpha \beta}^{(1,3)} \Bigr)\\
\bar{A}_{\alpha\beta}^{(1,2)} &= 8 \mu_{\beta}^2  \Bigl[\frac{35}{16} (12 \mu_{\alpha}^2 + 5 \mu_{\beta}^2)  -  
\frac{21}{8} (4 \mu_{\alpha}^2 + 5 \mu_{\beta}^2) \bar{\Xi}_{\alpha \beta}^{(1,2)} +
\frac{19}{4} \mu_{\beta}^2 \bar{\Xi}_{\alpha \beta}^{(1,3)}  
- \frac{1}{2} \mu_{\beta}^2 \bar{\Xi}_{\alpha \beta}^{(1,4)} 
+  7 \mu_{\alpha} \mu_{\beta} \bar{\Xi}_{\alpha \beta}^{(2,2)} -
2 \mu_{\alpha} \mu_{\beta} \bar{\Xi}_{\alpha \beta}^{(2,3)} 
\Bigr]\\
\nonumber
\bar{A}_{\alpha\beta}^{(2,2)} &= 8 \mu_{\beta}  \Bigl[\frac{35}{64} (40 \mu_{\alpha}^4 + 168 \mu_{\alpha}^2 \mu_{\beta}^2  +35 \mu_{\beta}^4)  -  
\frac{7}{8} \mu_{\beta}^2 (84 \mu_{\alpha}^2 + 35 \mu_{\beta}^2) \bar{\Xi}_{\alpha \beta}^{(1,2)} 
+  \frac{1}{8} \mu_{\beta}^2 (108 \mu_{\alpha}^2 + 133 \mu_{\beta}^2) \bar{\Xi}_{\alpha \beta}^{(1,3)}  \\
& - \frac{7}{2} \mu_{\beta}^4 \bar{\Xi}_{\alpha \beta}^{(1,4)}+
\frac{1}{4} \mu_{\beta}^4 \bar{\Xi}_{\alpha \beta}^{(1,5)}
+ \frac{7}{2} \mu_{\alpha} \mu_{\beta}  (4 \mu_{\alpha}^2 + 7 \mu_{\beta}^2) \bar{\Xi}_{\alpha \beta}^{(2,2)} - 
14 \mu_{\alpha} \mu_{\beta}^3 \bar{\Xi}_{\alpha \beta}^{(2,3)} +
2 \mu_{\alpha} \mu_{\beta}^3 \bar{\Xi}_{\alpha \beta}^{(2,4)} +
2 \mu_{\alpha}^2 \mu_{\beta}^2 \bar{\Xi}_{\alpha \beta}^{(3,3)}
\Bigr]\\
\bar{B}_{\alpha\beta}^{(0,0)} &= - 8 \mu_{\alpha}^{1/2} \mu_{\beta}^{1/2}  \\
\bar{B}_{\alpha\beta}^{(0,1)} &= - 8 \mu_{\alpha}^{3/2} \mu_{\beta}^{1/2}  \Bigl(  \frac{5}{2}   -  
\bar{\Xi}_{\alpha \beta}^{(1,2)} \Bigr)  \\
\bar{B}_{\alpha\beta}^{(1,1)} &= - 8 \mu_{\alpha}^{3/2} \mu_{\beta}^{3/2}  \Bigl(  \frac{55}{4}   -  
5  \bar{\Xi}_{\alpha \beta}^{(1,2)} +
 \bar{\Xi}_{\alpha \beta}^{(1,3)} -
2 \bar{\Xi}_{\alpha \beta}^{(2,2)} \Bigr)  \\
\bar{B}_{\alpha\beta}^{(0,2)} &=  - 4 \mu_{\alpha}^{5/2} \mu_{\beta}^{1/2}  \Bigl(  \frac{35}{4}   -  
7  \bar{\Xi}_{\alpha \beta}^{(1,2)} +
 \bar{\Xi}_{\alpha \beta}^{(1,3)}  \Bigr)  \\
\bar{B}_{\alpha\beta}^{(1,2)} &= - 8 \mu_{\alpha}^{5/2} \mu_{\beta}^{3/2}  \Bigl(   \frac{595}{16}   -  
\frac{189}{8}  \bar{\Xi}_{\alpha \beta}^{(1,2)} +
\frac{19}{4}  \bar{\Xi}_{\alpha \beta}^{(1,3)} -
\frac{1}{2}  \bar{\Xi}_{\alpha \beta}^{(1,4)} - 
7 \bar{\Xi}_{\alpha \beta}^{(2,2)} +
2 \bar{\Xi}_{\alpha \beta}^{(2,3)} \Bigr)  \\
\nonumber
\bar{B}_{\alpha\beta}^{(2,2)} &= - 8 \mu_{\alpha}^{5/2} \mu_{\beta}^{5/2}  \Bigl(   \frac{8505}{64}   -  
\frac{833}{8}  \bar{\Xi}_{\alpha \beta}^{(1,2)} +
\frac{241}{8}  \bar{\Xi}_{\alpha \beta}^{(1,3)} -
\frac{7}{2}  \bar{\Xi}_{\alpha \beta}^{(1,4)} +
\frac{1}{4}  \bar{\Xi}_{\alpha \beta}^{(1,5)}  
\label{eq: bracket-1-end}
- \frac{77}{2} \bar{\Xi}_{\alpha \beta}^{(2,2)} + 
14 \bar{\Xi}_{\alpha \beta}^{(2,3)} \\ 
& - 2 \bar{\Xi}_{\alpha \beta}^{(2,4)} +
2 \bar{\Xi}_{\alpha \beta}^{(3,3)}  \Bigr).
\end{align}

Elements of matrix $\tensor{L}$ are given by 
\begin{equation}
\label{eq: L-elem}
L_{\alpha\beta}^{(i,j)} = \frac{2}{5 T_i}
\Bigl( \delta_{\alpha\beta} \sum_{\chi =1}^{N} x_{\alpha} x_{\chi} C_{\alpha\chi}^{(i,j)} +
x_{\alpha} x_{\beta} K_{\alpha\beta}^{(i,j)} \Bigr),
\end{equation}
 where $C_{\alpha \beta}^{(l,k)} = 3  \nu_{\alpha \beta}/(16  n_{\beta}) \bar{C}_{\alpha \beta}^{(l,k)} $ and $K_{\alpha \beta}^{(l,k)} = 3 \nu_{\alpha \beta}/(16  n_{\beta}) \bar{K}_{\alpha \beta}^{(l,k)} $ and 
 \begin{align}
\label{eq: bracket-2}
\bar{C}_{\alpha\beta}^{(0,0)} &= \frac{16}{3} \mu_{\beta}  \Bigl( 
5 \mu_{\alpha}
+\frac{3}{2} \mu_{\beta} \bar{\Xi}_{\alpha \beta}^{(2,2)}
\Bigr)\\
\bar{C}_{\alpha\beta}^{(0,1)} &= \frac{16}{3} \mu_{\beta}^2  \Bigl(  
\frac{35}{2} \mu_{\alpha}    
- 7 \mu_{\alpha} \bar{\Xi}_{\alpha \beta}^{(1,2)} 
+\frac{21}{4} \mu_{\beta} \bar{\Xi}_{\alpha \beta}^{(2,2)}
-\frac{3}{2} \mu_{\beta} \bar{\Xi}_{\alpha \beta}^{(2,3)}
\Bigr)\\
\nonumber
\bar{C}_{\alpha\beta}^{(1,1)} &= \frac{16}{3} \mu_{\beta}  \Bigl[
\frac{1}{4} \mu_{\alpha} (140 \mu_{\alpha}^2 + 245 \mu_{\beta}^2)     
- 49 \mu_{\alpha} \mu_{\beta}^2 \bar{\Xi}_{\alpha \beta}^{(1,2)} 
+ 8  \mu_{\alpha} \mu_{\beta}^2 \bar{\Xi}_{\alpha \beta}^{(1,3)} 
+ \frac{1}{8} \mu_{\beta} (154 \mu_{\alpha}^2 + 147 \mu_{\beta}^2)   \bar{\Xi}_{\alpha \beta}^{(2,2)} \\
&- \frac{21}{2}  \mu_{\beta}^3 \bar{\Xi}_{\alpha \beta}^{(2,3)} 
+ \frac{3}{2}  \mu_{\beta}^3 \bar{\Xi}_{\alpha \beta}^{(2,4)} 
+ 3  \mu_{\alpha} \mu_{\beta}^2 \bar{\Xi}_{\alpha \beta}^{(3,3)} 
\Bigr]\\
\bar{K}_{\alpha\beta}^{(0,0)} &= -\frac{16}{3}  \mu_{\alpha} \mu_{\beta}  \Bigl( 
   5 
- \frac{3}{2} \bar{\Xi}_{\alpha \beta}^{(2,2)}
\Bigr)\\
\bar{K}_{\alpha\beta}^{(0,1)} &= \frac{16}{3} \mu_{\alpha}^2 \mu_{\beta}  \Bigl(  
-\frac{35}{2}
+ 7                      \bar{\Xi}_{\alpha \beta}^{(1,2)} 
+\frac{21}{4}       \bar{\Xi}_{\alpha \beta}^{(2,2)}
-\frac{3}{2}          \bar{\Xi}_{\alpha \beta}^{(2,3)}
\Bigr)\\
\label{eq: bracket-2-end}
\bar{K}_{\alpha\beta}^{(1,1)} &= -\frac{16}{3} \mu_{\alpha}^2 \mu_{\beta}^2  \Bigl(
\frac{385}{4}      
- 49                  \bar{\Xi}_{\alpha \beta}^{(1,2)} 
+ 8                   \bar{\Xi}_{\alpha \beta}^{(1,3)} 
- \frac{301}{8}  \bar{\Xi}_{\alpha \beta}^{(2,2)} 
+ \frac{21}{2} \bar{\Xi}_{\alpha \beta}^{(2,3)} 
- \frac{3}{2}      \bar{\Xi}_{\alpha \beta}^{(2,4)} 
+ 3                   \bar{\Xi}_{\alpha \beta}^{(3,3)} 
\Bigr).
\end{align}

In Eqs.~(\ref{eq: bracket-1})-(\ref{eq: bracket-1-end}) and (\ref{eq: bracket-2})-(\ref{eq: bracket-2-end}), $\mu_{\alpha} = m_{\alpha}/(m_{\alpha} + m_{\beta})$ and $\mu_{\beta} = m_{\beta}/(m_{\alpha} + m_{\beta})$ and due to symmetry properties of the bracket integrals $\bar{A}_{\alpha\beta}^{(i,j)} = \bar{A}_{\alpha\beta}^{(j,i)}$, $\bar{B}_{\alpha\beta}^{(i,j)} = \bar{B}_{\beta\alpha}^{(j,i)}$, $\bar{C}_{\alpha\beta}^{(i,j)} = \bar{C}_{\alpha\beta}^{(j,i)}$ and $\bar{K}_{\alpha\beta}^{(i,j)} = \bar{K}_{\beta\alpha}^{(j,i)}$~\cite{ ferziger1972mathematical}. The explicit expressions for the matrix elements provided in this Section can thus be used for evaluating the first to third order Chapman-Enskog approximations to the diffusive and heat fluxes and the first to second order Chapman-Enskog approximations to the viscosities. If higher accuracy is desired, one can retrieve the bracket integrals for larger $i$ and $j$ from prescription of Ref.~\cite{ ferziger1972mathematical}. However, as known from earlier works~\cite{braginskii1965transport,zhdanov2002transport}, going to higher order does not result in significant changes in the transport coefficients for weakly coupled plasmas and, as the more recent study~\cite{kagan2016influence}  revealed, the role of higher order corrections can only diminish with coupling within the EPT framework.


\endgroup

\end{widetext}

\section{Matrix elements in the weakly coupled limit}
\label{app: matrix-elem-weak}

In the weakly coupled limit~\cite{baalrud2014extending}
\begin{equation}
\label{eq: Coul-log-weak}
\bar{\Xi}_{\alpha \beta}^{(l,k)} = l (k-1)!,
\end{equation}
greatly simplifying Eqs.~(\ref{eq: bracket-1})-(\ref{eq: bracket-1-end}) and (\ref{eq: bracket-2})-(\ref{eq: bracket-2-end}) and making representation for the transport coefficients particularly compact:
\begingroup
\allowdisplaybreaks
\begin{align}
\label{eq: bracket-1-2-weak}
\bar{A}_{\alpha\beta}^{(0,0)} &= 8 \mu_{\beta}   \\
\bar{A}_{\alpha\beta}^{(0,1)} &= 12 \mu_{\beta}^2  \\
\bar{A}_{\alpha\beta}^{(1,1)} &= 2 \mu_{\beta} 
(30 \mu_{\alpha}^2 + 
16 \mu_{\alpha} \mu_{\beta} + 
13 \mu_{\beta}^2) \\
\bar{A}_{\alpha\beta}^{(0,2)} &= 15 \mu_{\beta}^3  \\
\bar{A}_{\alpha\beta}^{(1,2)} &= \frac{3}{2}  \mu_{\beta}^2  
(84 \mu_{\alpha}^2 + 
32 \mu_{\alpha} \mu_{\beta} + 
23 \mu_{\beta}^2)  \\
\nonumber
\bar{A}_{\alpha\beta}^{(2,2)} &= \frac{1}{8} \mu_{\beta}  
(1400 \mu_{\alpha}^4 + 
1792  \mu_{\alpha}^3 \mu_{\beta}     +
3672  \mu_{\alpha}^2 \mu_{\beta}^2 \\
&+ 1088  \mu_{\alpha}    \mu_{\beta}^3  + 
433 \mu_{\beta}^4)\\
\bar{B}_{\alpha\beta}^{(0,0)} &= - 8 \mu_{\alpha}^{1/2} \mu_{\beta}^{1/2}   \\
\bar{B}_{\alpha\beta}^{(0,1)} &= - 12 \mu_{\alpha}^{3/2} \mu_{\beta}^{1/2} \\
\bar{B}_{\alpha\beta}^{(1,1)} &= - 54 \mu_{\alpha}^{3/2} \mu_{\beta}^{3/2} \\
\bar{B}_{\alpha\beta}^{(0,2)} &=  - 15 \mu_{\alpha}^{5/2} \mu_{\beta}^{1/2} \\
\bar{B}_{\alpha\beta}^{(1,2)} &= - \frac{225}{2} \mu_{\alpha}^{5/2} \mu_{\beta}^{3/2}  \\
\bar{B}_{\alpha\beta}^{(2,2)} &= - \frac{2625}{8} \mu_{\alpha}^{5/2} \mu_{\beta}^{5/2} \\
\bar{C}_{\alpha\beta}^{(0,0)} &= \frac{16}{3} \mu_{\beta} ( 5 \mu_{\alpha} + 3 \mu_{\beta})   \\
\bar{C}_{\alpha\beta}^{(0,1)} &= 8 \mu_{\beta}^2 ( 7 \mu_{\alpha} + 3 \mu_{\beta}) \\
\bar{C}_{\alpha\beta}^{(1,1)} &= \frac{4}{3} \mu_{\beta} 
(140 \mu_{\alpha}^3 + 
154 \mu_{\alpha}^2 \mu_{\beta} + 
185 \mu_{\alpha} \mu_{\beta}^2 + 
51  \mu_{\beta}^3) \\
\bar{K}_{\alpha\beta}^{(0,0)} &= - \frac{32}{3} \mu_{\alpha} \mu_{\beta}   \\
\bar{K}_{\alpha\beta}^{(0,1)} &= - 32               \mu_{\alpha}^2 \mu_{\beta} \\
\bar{K}_{\alpha\beta}^{(1,1)} &= - 160 \mu_{\alpha}^2 \mu_{\beta}^2 .
\end{align}
The matrix elements $M_{\alpha\beta}^{(i,j)}$ and $L_{\alpha\beta}^{(i,j)}$ are obtained by using the above expressions in Eqs.~(\ref{eq: M-elem-2}) and (\ref{eq: L-elem}), respectively.
\endgroup

\section{Transport in weakly coupled plasmas }
\label{app: diff-weak}

Here we illustrate application of the presented formulary by considering transport in weakly coupled plasmas. We begin by noticing that $p_{\alpha} = x_{\alpha} p_i$ and so the first term on the right side of Eq.~(\ref{eq: diff-force}) becomes
\begin{equation}
\label{eq: press-grad}
\frac{\nabla p_{\alpha} - c_{\alpha} \nabla p_i}{p_i} = 
\nabla x_{\alpha} + (x_{\alpha} - c_{\alpha} ) \nabla \ln{p_i}.
\end{equation}

Next, we evaluate the second term on the right side of  Eq.~(\ref{eq: diff-force}) with the help of Eq.~(\ref{eq: ext-force}). To do so we first notice that $\vec{ R}_{ei}^{(T)} = - B_e^{(e)} n_e \nabla T_e = - \sum_{\beta =1}^{N}  \vec{ R}_{\beta e}^{(T)}$, where $\vec{ R}_{ei}^{(T)}$ is the total thermal force exerted on electrons by all the ion species. We also use that contribution to this total force from an ion species $\beta$ is proportional to $n_{\beta} Z_{\beta}^2$, making
\begin{equation}
\label{eq: ion-el-thermal-force}
\vec{ R}_{\beta e}^{(T)} = 
\frac{n_{\beta} Z_{\beta}^2}{Z_{\bf eff}} B_e^{(e)} \nabla T_e,
\end{equation}
where $Z_{\bf eff} \equiv ( \sum_{\beta =1}^{N} n_{\beta} Z_{\beta}^2) /n_e $ is the
effective ion charge number and the electron-ion thermal force coefficient can be found in literature~\cite{zhdanov2002transport}:
\begin{equation}
\label{eq: el-thermal}
B_e^{(e)}  \approx 
\frac{0.47+0.94Z_{\bf eff}^{-1}}{0.31+1.20Z_{\bf eff}^{-1}+0.41Z_{\bf eff}^{-2}}.                    
\end{equation}

As a proof of principle one can also obtain electron-ion thermal force  from formulas of subsection~\ref{app: diff}     by taking $N=2$ and interpreting component  ``1" as the electrons and component ``2"   as the single ion species with the charge number $Z_{\bf eff}$. Then, by setting $n_2 = Z_{\bf eff} n_1$ to enforce  quasi-neutrality and evaluating $B_e^{(e)} =  c_1 n D_{1}^{(T)}/(n_1 D_{12})$~\footnote{E.g. see Eq.~(5) of Ref.~\cite{kagan2016influence} }
  with $Z_{\bf eff}$ equal to $1,2,3$ and 4 one can reproduce Braginskii's results for the thermal force coefficient~\cite{braginskii1965transport}. By utilizing $(m_e/m_i)^{1/2} \ll 1$ one can also simplify the determinant on the right side of Eq.~(\ref{eq: thermo-diff}) to recover the analytical expression~(\ref{eq: el-thermal}), which was originally obtained by Zhdanov by separating the electron and ion moment equations~\cite{zhdanov2002transport}. Of course, Eq.~(\ref{eq: el-thermal}) evaluated at $Z_{\bf eff} = 1,2,3$ and 4  gives Braginskii's results as well.

Using Eq.~(\ref{eq: press-grad}) and   Eq.~(\ref{eq: ext-force}) along with Eq.~(\ref{eq: ion-el-thermal-force}), we find the diffusion driving force from Eq.~(\ref{eq: diff-force})  
\begin{multline}
\label{eq: ext-force-contrib}
\vec{d}_{\alpha} = \nabla x_{\alpha} + (x_{\alpha} - c_{\alpha} ) \nabla \ln{p_i} -
c_{\alpha} \Bigl( \frac{ Z_{\alpha}^2}{m_{\alpha}} - \sum_{\beta = 1}^N  \frac{ c_{\beta} Z_{\beta}^2}{m_{\beta}}  \Bigr) \times \\
\frac{\rho}{n_i} \frac{T_e}{T_i}
\frac{B_e^{(e)}}{Z_{\bf eff}} \nabla \log T_e 
- c_{\alpha} \Bigl( \frac{ Z_{\alpha}}{m_{\alpha}} - \sum_{\beta = 1}^N  \frac{ c_{\beta} Z_{\beta}}{m_{\beta}}  \Bigr)
\frac{\rho}{n_i} \frac{ e \vec{E}}{T_i}.
\end{multline}

Eq.~(\ref{eq: ext-force-contrib}) and formulas of Section~\ref{app: formulary} with the matrix elements of Section~\ref{app: matrix-elem-weak} give fully analytical expressions for  all the transport coefficients in a weakly coupled plasma with $N$ ion species. We now consider the case $N=2$ and write the  diffusive \emph{mass} flux $\rho_{\alpha} \vec{V}_{\alpha}$ of the lighter ion species in the Landau-Lifshitz form~\cite{landau1987fluid}
\begin{multline}
\label{eq: LL-flux}
\vec{i} =
- \rho D \Bigl( \nabla c +k_p \nabla \log{p_i} + \frac{e k_E}{T_i}\nabla \Phi + k_T^{(i)} \nabla \log{T_i} \\
 + k_T^{(e)} \nabla \log{T_e}\Bigr),
\end{multline}
where $\vec{i} \equiv \vec{i}_l = - \vec{i}_h$ and $c \equiv c_l = 1- c_h$ with subscripts ``$l$" and ``$h$" denoting the light and heavy ion species, respectively. Using that 
\begin{equation}
\label{eq: grad-x}
\nabla x_l = - \nabla x_h = \frac{1}{m_l m_h} \Bigl(\frac{\rho}{n_i} \Bigr)^2 \nabla c
\end{equation}
and comparing Eq.~(\ref{eq: LL-flux}) and Eqs.~(\ref{eq: diff-flux-gen})-(\ref{eq: ext-force}) it is straightforward to see that the ordinary diffusion coefficient given by Eq.~(\ref{eq: ordin-diff}) is related to the classical diffusion coefficient $D$ of Landau and Lifzhitz through
\begin{equation}
\label{eq: classic-diff}
D = - \frac{ \rho^2 D_{lh}}{m_l m_h n_i^2},
\end{equation}
and is also equal to the ``binary diffusion coefficient" $\mathcal{D}_{12}$ of Ferziger and Kaper. In turn, the thermo-diffusion ratio with the ion temperature gradient can be recovered from
\begin{equation}
\label{eq: thermo-diff-LL}
k_T^{(i)} = \frac{c m_l m_h n_i^2 D_{l}^{(T)}}{  \rho^2 D_{lh}}.
\end{equation}

One can then utilize Eqs.~(\ref{eq: ordin-diff}) and  (\ref{eq: thermo-diff}) to reproduce the results for the thermo-diffusion ratio with the ion temperature gradient from Fig. 2 of our earlier work~\cite{kagan2014thermo}. The dynamic friction results, shown in Fig. 1 of the same publication,  can be reproduced by computing $A_{lh} = [D_{lh}]_1/[D_{lh}]_3$ from Eq.~(\ref{eq: ordin-diff}). 

Finally, our earlier result for the baro-diffusion ratio $k_p$ can be reproduced by evaluating the right side of Eq.~(\ref{eq: press-grad}) with the help of Eq.~(\ref{eq: grad-x}), and the results for the electro-diffusion ratio $k_E$ and  thermo-diffusion ratio with the electron temperature gradient $k_T^{(e)}$ by evaluating the right side of Eq.~(\ref{eq: ext-force-contrib}) with $N=2$.

\section{Numerical routines }
\label{sec: routines}

The presented expressions for the transport coefficients have been implemented in Matlab routines, which can be downloaded from the corresponding links in the ``ancillary files" field on this article webpage. These routines do not involve symbolic math operations and can be readily adopted in any other standard programming environment. 

For coupled plasmas one has to use formulas of Section~\ref{app: formulary} along with matrix elements of Section~\ref{app: matrix-elem}. These are implemented in ``\textbf{heat\underline{~}diffusion.m}" and ``\textbf{viscosity.m}", which return the corresponding transport coefficients along with the reduced masses $\mu_{\alpha\beta}$ and effective binary collision frequencies $\nu_{\alpha\beta}$. The arguments are the temperature, $N$-element vectors of the species masses, charge numbers and number densities, and  an array of $\Xi_{\alpha \beta}^{(l,k)}$. 

The generalized Coulomb logarithms $\Xi_{\alpha \beta}^{(l,k)}$ are to be calculated separately as described in Ref.~\cite{baalrud2013effective}. The Matlab data file ``\textbf{xi\underline{~}lk\underline{~}ocp.mat}" with the generalized Coulomb logarithms for the one-component plasma (OCP) of hydrogen is also included. Since the effective potential introduced in Ref.~\cite{baalrud2013effective} is not sensitive to relative concentrations of species with the same charge number, one can use the same data for evaluating transport in a mixture of hydrogen isotopes as well as in OCP. To illustrate application of the transport formalism for coupled plasmas, we include routine ``\textbf{example\underline{~}coupled.m}", which submits the data from  \textbf{xi\underline{~}lk\underline{~}ocp.mat} to \textbf{heat\underline{~}diffusion.m} and \textbf{viscosity.m} to evaluate the transport in OCP and in the binary DT mixture. The output is then plotted to reproduce the OCP results shown in Fig.~7 of Ref.~\cite{kagan2016influence} and the DT results shown in Fig.~1 of the same publication.

To calculate the transport coefficients for weakly coupled plasmas one can use the same routines by submitting the right side of Eq.~(\ref{eq: Coul-log-weak}) for $\bar{\Xi}_{\alpha \beta}^{(l,k)}$. However, it is  more computationally efficient to directly code the much simplified expressions for the bracket integrals from Section~\ref{app: matrix-elem-weak} of this note. The simplified routines 
``\textbf{heat\underline{~}diffusion\underline{~}weak.m}" and 
``\textbf{viscosity\underline{~}weak.m}" are also included as ancillary files. The Matlab script ``\textbf{example\underline{~}weak.m}" illustrates their application by using \textbf{heat\underline{~}diffusion\underline{~}weak.m} to calculate the dynamic friction coefficient $A_{lh}$ and thermo-diffusion ratio $k_T^{(i)}$ as functions of $c$ for a binary ionic mixture. By appropriately setting the  ion masses and charge numbers one can then reproduce the results of Figs.~1 and 2  of Ref.~\cite{kagan2014thermo}.

\acknowledgements
The authors would like to thank J. Daligault of LANL and  A.A. Stepanenko and V.M. Zhdanov of MEPhI for many useful discussions. This work was partially supported by the ASC Thermonuclear Burn Initiative under the auspices of the U.S. Dept. of Energy by the Los Alamos National Security, LLC, Los Alamos National Laboratory under Contract No. DE-AC52-06NA25396.


\begin{thebibliography}{10}
\expandafter\ifx\csname url\endcsname\relax
  \def\url#1{{\tt #1}}\fi
\expandafter\ifx\csname urlprefix\endcsname\relax\def\urlprefix{URL }\fi
\providecommand{\eprint}[2][]{\url{#2}}

\bibitem{baalrud2013effective}
Baalrud S~D and Daligault J 2013 {\em Physical Review Letters\/} {\bf 110}
  235001

\bibitem{zhdanov2002transport}
Zhdanov V~M 2002 {\em Transport Processes in Multicomponent Plasma\/} (CRC
  Press)

\bibitem{hirschfelder1954molecular}
Hirschfelder J~O, Curtiss C~F and Bird R~B 1954 {\em Molecular theory of gases
  and liquids\/} (Wiley New York)

\bibitem{devoto1966transport}
Devoto R 1966 {\em Physics of Fluids\/} {\bf 9} 1230--1240

\bibitem{ferziger1972mathematical}
Ferziger J~H and Kaper H~G 1972 {\em Mathematical theory of transport processes
  in gases\/} (North Holland)

\bibitem{baalrud2014extending}
Baalrud S~D and Daligault J 2014 {\em Physics of Plasmas\/} {\bf 21} 055707

\bibitem{kagan2012electro}
Kagan G and Tang X~Z 2012 {\em Physics of Plasmas\/} {\bf 19} 082709

\bibitem{braginskii1965transport}
Braginskii S 1965 {\em Reviews of Plasma Physics\/} {\bf 1} 205

\bibitem{kagan2016influence}
Kagan G, Baalrud S~D and Daligault J 2016 {\em \emph{``Influence of coupling on
  thermal forces and dynamic friction in plasmas with multiple ion species"},
  arXiv:1609.00742\/}

\bibitem{Note1}
E.g. see Eq.~(5) of Ref.~\cite {kagan2016influence}

\bibitem{landau1987fluid}
Landau L and Lifshitz E 1987 {\em Fluid Mechanics\/} (Oxford: Pergamon Press)

\bibitem{kagan2014thermo}
Kagan G and Tang X~Z 2014 {\em Physics Letters A\/} {\bf 378} 1531--1535

\end{thebibliography}

\providecommand{\newblock}{}

\end{document}